\documentclass[12pt]{article}
\usepackage{epsf}
\usepackage{graphicx}
\usepackage{lscape}
\usepackage{amssymb}
\usepackage{pazh}

\tightenlines

\def\пм{$\pm$}
\def\*{$^{*}$}

\def\a{$^{\mbox{\small a}}$}
\def\b{$^{\mbox{\small b}}$}

\def\ИН{\mbox{\<<ИНТЕГРАЛ\>>}}
\def\4у{\mbox{4U 0115+63}}
\def\В03{\mbox{!!??XX??!!}}

\def\ергс{эрг~с$^{-1}$}
\def\ергсм{эрг~см$^{-2}$~с$^{-1}$}
\def\etal{{\it et~al.}}

\begin{document}

{\footnotesize Astronomy Letters, Vol. 33, No. 6, 2007, pp. 368-384.
Translated from Pis'ma v Astronomicheskii Zhurnal, Vol. 33, No. 6,
2007, pp. 417-434.}

\title{\bf 4U 0115+63 from RXTE and INTEGRAL Data: Pulse Profile and Cyclotron Line Energy}

\author{\bf S. S. Tsygankov$^{1,2 \,*}$, A. A. Lutovinov$^{1,2}$, E. M. Churazov$^{2,1}$ and R. A. Sunyaev$^{2,1}$}

\affil{
$^{1}$Space Research Institute, Profsoyuznaya str. 84/32, Moscow 117997, Russia\\
$^{2}$MPI for Astrophysik, Karl-Schwarzschild str. 1, Garching, 85741, Germany}

\sloppypar
\vspace{2mm}
\noindent


We analyze the observations of the transient X-ray pulsar 4U 0115+63 with the RXTE and
INTEGRAL observatories in a wide X-ray (3--100 keV) energy band during its intense outbursts in 1999
and 2004. The energy of the fundamental harmonic of the cyclotron resonance absorption line near the
maximum of the X-ray flux from the source (luminosity range  $5\times10^{37}$ -- $2\times10^{38}$ erg s$^{-1}$) is $\sim11$ keV.
When the pulsar luminosity falls below $\sim5\times10^{37}$ erg s$^{-1}$, the energy of the fundamental harmonic is
displaced sharply toward the high energies, up to $\sim16$ keV. Under the assumption of a dipole magnetic
field configuration, this change in cyclotron harmonic energy corresponds to a decrease in the height of
the emitting region by $\sim2$ km, while other spectral parameters, in particular, the cutoff energy, remain
essentially constant. At a luminosity $\sim7\times10^{37}$ erg s$^{-1}$ , four almost equidistant cyclotron line harmonics
are clearly seen in the spectrum. This suggests that either the region where the emission originates is
compact or the emergent spectrum from different (in height) segments of the accretion column is uniform.
We have found significant pulse profile variations with energy, luminosity, and time. In particular, we show
that the profile variations from pulse to pulse are not reduced to a simple modulation of the accretion rate
specified by external conditions.


\vfill

{$^{*}$ E-mail: st@hea.iki.rssi.ru}
\newpage
\thispagestyle{empty}
\setcounter{page}{1}

\section*{INTRODUCTION}

   The X-ray source 4U 0115+63 was discovered
by the UHURU observatory more than 30 years ago
(Giacconi et al. 1972; Forman et al. 1978). During the
SAS-3 observations in 1978, Rappaport et al. (1978)
determined the binary's main parameters based on
the known pulsation period of $\sim3.6$ s (Cominsky et al.
1978): orbital period $\sim24.3$ days, orbital eccentricity 0.34,
and projected semimajor axis of the relativistic object
$a_{x}$sin$i\sim140$ light seconds (see also Tamura et al.
(1992) and Lutovinov et al. (1994) for an improvement of the 
parameters). The optical observations of
the star V635 Cas (Hutchings and Crampton 1981;
Kholopov et al. 1981), the normal companion of the
X-ray source 4U 0115+63, performed by Negueruela
and Okazaki (2001) allowed the star's spectral type
to be improved, B0.2Ve, and the binary's distance to
be estimated, 7--8 kpc.

In addition to its transient behavior (which ensures
a wide dynamic range of the object's observed luminosities), 
the X-ray pulsar 4U 0115+63 is unique
in its spectral characteristics. The cyclotron resonance 
absorption line was first detected in its 
spectrum almost 30 years ago. At present, it is the only
object in the spectrum of which five cyclotron line
harmonics were detected. The properties of the 
cyclotron feature in the source under consideration were
studied in detail using data from many observatories
(Wheaton et al. 1979; White et al. 1983; Mihara et al.
1998, 2004; Heindl et al. 1999; Santangelo et al.
1999; Lutovinov et al. 2000). It was shown that the
source spectrum is subject to significant variations
on a time scale shorter than the spin period of the
neutron star and that different cyclotron frequency
harmonics behave differently depending on the pulse
phase; in particular, the third harmonic is present only
on the descent of the second (smaller) peak of the
pulse profile (Heindl et al. 1999). Mihara et al. (1998)
found that only one cyclotron absorption harmonic
at $\sim 16$ keV was detected in the source spectrum
during its observations in 1991 instead of the two
lines at $\sim 12$ and $\sim 22$ keV observed in 1990. Using the
RXTE data obtained during another intense outburst
in March­April 1999, Nakajima et al. (2006) 
confirmed that the cyclotron line position in the source
spectrum depends on the pulsar luminosity, while the
single line at $\sim 16$ keV is most likely the "upward"-displaced fundamental harmonic.

 Despite many observations and studies of the X-ray 
pulsar 4U 0115+63, a number of poorly studied
questions related to the variations in the source pulse
profile and its spectral characteristics with source
intensity and energy band still remain.

   In this paper, we perform a timing analysis of the
pulsar emission on the scale of the pulsation period
and investigate its spectral properties as a function of
the intrinsic luminosity using the RXTE and INTEGRAL 
data obtained during the 2004 outburst and
RXTE data for the previous 1999 outburst.

\section*{OBSERVATIONS AND DATA ANALYSIS}

    In this paper, we use the observations of the X-ray 
pulsar 4U 0115+63 during its intense outbursts
in March­April 1999 and September­October 2004
with the instruments of the RXTE orbital astrophysical 
observatory (Bradt et al. 1993) -- the PCA
and HEXTE spectrometers (observations 
40051-05-XX-XX, 40070-01-XX-00, 40411-01-XX-00,
90014-02-XX-XX, and 90089-01-XX-XX) operating 
in the $3-20$ and $15-250$ keV energy bands,
respectively. The PCA spectrometer is a system of
five proportional xenon/propane counters with an
effective area of $\sim$6400 cm$^2$ at 6--7 keV and an energy
resolution of $\sim$18\%. The HEXTE spectrometers
is a system of two independent packages of four
NaI(Tl)/CsI(Na) phoswich detectors swinging with
an interval of 16 s for the observation of background
areas at a distance of 1.5$^\circ$ from the source. At each
specific time, the source can be observed only by one
of the two detector packages; thus, the effective area
of the HEXTE detectors is $\sim$700 cm$^2$ . The standard
FTOOLS/LHEASOFT 5.3 software package was
used to process the RXTE data.

   To study in detail the properties of the hard
($>20$ keV) X-ray emission from the pulsar, we used
the data from the IBIS telescope (ISGRI detector;
Lebrun et al. 2003) of the INTEGRAL observatory (Winkler et al. 2003) 
obtained during its outburst that began late in August 2004 (Lutovinov
et al. 2004). We processed the IBIS data for our
timing analysis using the software developed and
maintained at the National Astrophysical Institute
in Palermo, Italy (http://www.pa.iasf.cnr.it/\~ferrigno/INTEGRALsoftware.html); 
a description of
the data processing technique can be found in Mineo
et al. (2006). For our spectral analysis, we used
the software and methods developed at the Space
Research Institute of the Russian Academy of 
Sciences (Moscow, Russia) and described in Revnivtsev
et al. (2004). Analysis of a large set of calibration
observations for the Crab Nebula showed that the
uncertainties in the derived source energy spectra
related to the evolution of the detector background
and its characteristics do not exceed $\sim3$\%. The latter
was added as a systematic uncertainty during our
spectral analysis with the XSPEC software package.
For our spectral analysis at energies below 20 keV,
we used JEM-X data (Lund et al. 2003), which
were processed with the standard OSA 5.1 package
provided by the INTEGRAL Science Data Center
(http://isdc.unige.ch).

   Tables 1 and 2 list the RXTE and INTEGRAL
observations of the X-ray pulsar 4U 0115+63 in 1999
and 2004. Column 1 gives the observation date and
number; columns 2 and 3 provide the flux from the
source and its bolometric luminosity (calculated for
an assumed distance to the source of 7 kpc) in the
range of main energy release $3-100$ keV, respectively.
The typical RXTE exposures times lie within the
range from $\sim0.3$ to $\sim4$ ks; the INTEGRAL exposure
time for the source is $\sim200$ ks. All of the measurement
errors given here correspond to $1\sigma$.

\newpage


\begin{table}[p]
\noindent
\centering

{\bf Table 1. }{RXTE observations of the pulsar 4U 0115+63 in 1999}\\
\centering
\vspace{1mm}
\footnotesize{
\begin{tabular}{c|c|c}
\hline\hline
 Date, MJD &   Flux\a,                      &  Luminosity\b,   \\
(pointing)  &   $\times10^{-9}$ erg s$^{-1}$ cm$^{-2}$  &  $10^{37}$ erg s$^{-1}$   \\
\hline
51240.15 (40411-01-01-00) & $ 12.2 \pm 1.1 $ & $ 7.2 \pm 0.6 $ \\
51241.56 (40411-01-02-00) & $ 12.3 \pm 1.0 $ & $ 7.3 \pm 0.6 $ \\
51242.87 (40411-01-03-00) & $ 16.5 \pm 0.9 $ & $ 9.7 \pm 0.5 $ \\
51243.87 (40411-01-04-00) & $ 19.1 \pm 1.2 $ & $ 11.2 \pm 0.7 $ \\
51244.31 (40070-01-01-00) & $ 20.2 \pm 0.1 $ & $ 11.9 \pm 0.1 $ \\
51244.87 (40411-01-05-00) & $ 20.1 \pm 1.3 $ & $ 11.8 \pm 0.8 $ \\
51246.27 (40411-01-06-00) & $ 22.1 \pm 1.7 $ & $ 13.0 \pm 1.0 $ \\
51246.83 (40411-01-07-00) & $ 26.4 \pm 2.0 $ & $ 15.5 \pm 1.1 $ \\
51247.76 (40411-01-08-00) & $ 21.4 \pm 1.3 $ & $ 12.6 \pm 0.8 $ \\
51248.38 (40411-01-09-00) & $ 21.9 \pm 1.4 $ & $ 12.9 \pm 0.8 $ \\
51248.94 (40070-01-02-00) & $ 24.4 \pm 1.1 $ & $ 14.4 \pm 0.6 $ \\
51249.16 (40070-01-03-00) & $ 24.9 \pm 0.3 $ & $ 14.6 \pm 0.2 $ \\
51249.76 (40411-01-10-00) & $ 22.5 \pm 1.4 $ & $ 13.2 \pm 0.8 $ \\
51250.80 (40411-01-11-00) & $ 24.6 \pm 1.3 $ & $ 14.5 \pm 0.8 $ \\
51251.28 (40411-01-12-00) & $ 23.5 \pm 1.4 $ & $ 13.8 \pm 0.8 $ \\
51252.34 (40411-01-13-00) & $ 20.8 \pm 1.5 $ & $ 12.3 \pm 0.9 $ \\
51253.40 (40411-01-14-00) & $ 21.0 \pm 1.1 $ & $ 12.3 \pm 0.6 $ \\
51254.61 (40411-01-15-00) & $ 23.7 \pm 1.4 $ & $ 13.9 \pm 0.8 $ \\
51255.33 (40411-01-16-00) & $ 23.2 \pm 1.5 $ & $ 13.6 \pm 0.9 $ \\
51256.22 (40051-05-01-00) & $ 21.0 \pm 0.8 $ & $ 12.3 \pm 0.4 $ \\
51256.54 (40411-01-17-00) & $ 19.0 \pm 0.8 $ & $ 11.2 \pm 0.5 $ \\
51257.16 (40411-01-18-00) & $ 18.8 \pm 1.2 $ & $ 11.1 \pm 0.7 $ \\
51258.31 (40051-05-02-00) & $ 20.9 \pm 0.9 $ & $ 12.3 \pm 0.5 $ \\
51258.55 (40411-01-19-00) & $ 19.8 \pm 0.7 $ & $ 11.6 \pm 0.4 $ \\
51259.20 (40411-01-20-00) & $ 16.5 \pm 0.8 $ & $ 9.7 \pm 0.4 $ \\
51260.29 (40051-05-03-00) & $ 18.6 \pm 1.0 $ & $ 10.9 \pm 0.6 $ \\
51260.42 (40411-01-21-00) & $ 18.7 \pm 2.0 $ & $ 11.0 \pm 1.1 $ \\
51261.54 (40411-01-22-00) & $ 17.0 \pm 1.1 $ & $ 10.0 \pm 0.7 $ \\
51262.23 (40051-05-04-00) & $ 13.7 \pm 0.4 $ & $ 8.1 \pm 0.2 $ \\
51263.25 (40411-01-23-00) & $ 12.6 \pm 1.2 $ & $ 7.4 \pm 0.7 $ \\
51264.39 (40051-05-05-00) & $ 13.6 \pm 0.7 $ & $ 8.0 \pm 0.4 $ \\
51265.19 (40070-01-04-00) & $ 12.8 \pm 0.1 $ & $ 7.5 \pm 0.1 $ \\
51266.07 (40051-05-06-00) & $ 11.6 \pm 0.3 $ & $ 6.8 \pm 0.2 $ \\
51266.32 (40070-01-05-00) & $ 11.3 \pm 0.1 $ & $ 6.6 \pm 0.1 $ \\
51268.22 (40051-05-07-00) & $ 9.4 \pm 0.5 $ & $ 5.5 \pm 0.3 $ \\
51269.16 (40051-05-15-02) & $ 8.8 \pm 0.3 $ & $ 5.2 \pm 0.2 $ \\
51270.22 (40051-05-08-00) & $ 8.3 \pm 0.1 $ & $ 4.9 \pm 0.1 $ \\
51271.11 (40051-05-15-01) & $ 8.1 \pm 0.7 $ & $ 4.7 \pm 0.4 $ \\
51272.21 (40051-05-09-00) & $ 7.0 \pm 0.1 $ & $ 4.1 \pm 0.1 $ \\
51273.47 (40051-05-15-00) & $ 6.0 \pm 1.0 $ & $ 3.5 \pm 0.6 $ \\
51274.14 (40051-05-10-00) & $ 5.9 \pm 0.1 $ & $ 3.5 \pm 0.1 $ \\
51276.09 (40051-05-11-00) & $ 5.0 \pm 0.3 $ & $ 2.9 \pm 0.2 $ \\
51278.21 (40051-05-12-00) & $ 3.9 \pm 0.1 $ & $ 2.3 \pm 0.1 $ \\
51280.14 (40051-05-13-00) & $ 2.5 \pm 0.2 $ & $ 1.5 \pm 0.1 $ \\
51282.09 (40051-05-14-00) & $ 2.0 \pm 0.1 $ & $ 1.2 \pm 0.1 $ \\
51284.92 (40411-01-24-00) & $ 3.4 \pm 0.2 $ & $ 2.0 \pm 0.1 $ \\
51286.92 (40411-01-25-00) & $ 2.3 \pm 0.2 $ & $ 1.4 \pm 0.1 $ \\
51288.06 (40411-01-26-00) & $ 0.2 \pm 0.3 $ & $ 0.1 \pm 0.1 $ \\

\hline
\end{tabular}
\vspace{3mm}

\begin{tabular}{ll}
\a & In the 3­100 keV energy band. \\
\b & In the 3­100 keV energy band for an assumed source distance
of $d=7$ kpc.\\
\end{tabular}}
\end{table}




\begin{table}[p]
\noindent
\centering

{\bf Table 2. }{ RXTE and INTEGRAL observations of the pulsar 4U 0115+63 in 2004}\\
\centering
\vspace{1mm}
\footnotesize{
\begin{tabular}{c|c|c}
\hline\hline
 Date, MJD &   Flux\a,                      &  Luminosity\b,   \\
(pointing)  &   $\times10^{-9}$ erg s$^{-1}$ cm$^{-2}$  &  $10^{37}$ erg s$^{-1}$   \\
\hline
\multicolumn{3}{c}{}\\[-4mm]
\multicolumn{3}{c}{From RXTE (PCA and HEXTE) data}\\[2mm]
\hline
&&\\ [-4mm]
53260.18 (90089-01-03-01) & $  18.3 \pm  2.4  $ & $ 10.8  \pm 1.4 $ \\
53262.20 (90089-01-03-00) & $  22.8 \pm  2.4  $ & $ 13.4 \pm  1.4 $ \\
53265.08 (90089-01-04-06) & $  18.7 \pm  0.6 $ & $  11.0 \pm  0.4$ \\
53267.24 (90089-01-04-04) & $  16.8  \pm 0.5 $ & $  9.9 \pm  0.3$ \\
53269.13 (90089-01-04-02) & $  17.2 \pm  0.4 $ & $  10.1 \pm  0.2$ \\
53270.51 (90089-01-04-00) & $  16.2 \pm  0.6 $ & $  9.5 \pm  0.4$ \\
53271.97 (90089-01-04-03) & $  14.7 \pm  0.1 $ & $  8.6  \pm 0.1$ \\
53272.82 (90089-01-05-00) & $  13.9 \pm  0.3 $ & $  8.2 \pm  0.2$ \\
53272.96 (90089-01-05-05) & $  13.8 \pm  0.4 $ & $  8.1 \pm  0.2$ \\
53274.85 (90089-01-05-01) & $  12.6 \pm  1.4 $ & $  7.4  \pm 0.8$ \\
53275.69 (90089-01-05-02) & $  11.9 \pm  0.6 $ & $  7.0  \pm 0.4$ \\
53276.75 (90089-01-05-03) & $  11.0 \pm  0.2 $ & $  6.5  \pm 0.1$ \\
53278.65 (90089-01-05-04) & $  9.6 \pm  0.4 $ & $  5.6  \pm 0.2$ \\
53280.73 (90089-01-06-00) & $  8.0 \pm  1.0 $ & $  4.7  \pm 0.6$ \\
53282.65 (90014-02-01-00) & $  6.8 \pm  0.1 $ & $  4.0  \pm 0.1$ \\
53284.62 (90014-02-01-02) & $  5.0 \pm  0.1 $ & $  2.9  \pm 0.1$ \\
53285.99 (90014-02-01-01) & $  3.8 \pm  0.4 $ & $  2.2 \pm  0.2$ \\
53287.83 (90014-02-02-00) & $  3.2 \pm  0.4 $ & $  1.9 \pm  0.2$ \\
53289.54 (90014-02-02-01) & $  2.5 \pm  0.3  $ & $ 1.5 \pm  0.2$ \\
53290.65 (90014-02-02-02) & $  1.9 \pm  0.1  $ & $ 1.1 \pm  0.1$ \\
53291.83 (90014-02-02-03) & $  1.5 \pm  1.0 $ & $  0.9  \pm 0.6$ \\
53293.46 (90014-02-03-00) & $  1.1 \pm  0.1 $ & $  0.6  \pm 0.1$ \\
53295.70 (90014-02-03-01) & $  0.4 \pm  0.5 $ & $  0.2  \pm 0.3$ \\
\hline
\multicolumn{3}{c}{}\\[-4mm]
\multicolumn{3}{c}{From INTEGRAL (JEM-X and IBIS) data}\\[2mm]
\hline
&&\\ [-4mm]
53273.8 (238 орбита) & $  12.5 \pm  1.3 $ & $  7.4  \pm 0.8$ \\

\hline
\end{tabular}
\vspace{3mm}

\begin{tabular}{ll}
\a & In the 3­100 keV energy band. \\
\b & In the 3­100 keV energy band for an assumed source distance
of $d=7$ kpc.\\
\end{tabular}}
\end{table}


\section*{THE PULSE PROFILE}

{\it The Average Pulse Profile, its Evolution with Luminosity and Energy}

    We analyzed the properties of the pulse profile 
using the data obtained during the 1999 outburst, since
it had a wider luminosity range and was observed
with the RXTE instruments from an earlier phase.
The results of our analysis of the more recent 
observations in 2004 showed no significant differences
from those obtained for the 1999 outburst. Assuming
the distance to the binary to be 7 kpc, the range of
observed luminosities for the source during its 
outburst was $\sim$ (1--15) $\times 10^{37}$ erg s$^{-1}$ . Due to such a
high intensity of the pulsar emission, we were able not
only to trace the dependence of the pulse profile shape
on the object's intrinsic luminosity in various energy
bands with a good statistical significance, but also to
investigate its variability on the scale of the pulsation
period.

    Figure 1 shows the background-corrected phase
light curves of the pulsar 4U 0115+63 at various
source luminosities in various energy bands obtained
from RXTE data. In the figure, the columns are 
arranged from the outburst onset from left to right
and correspond to luminosities $\sim 7.3\times10^{37}$, $\sim
14.6\times10^{37}$, $\sim 6.6\times10^{37}$ and $\sim
1.5\times10^{37}$ erg s$^{-1}$. The first
and third columns reflect the evolution of the pulse
profile with energy for the rise and decay of the 
outburst, respectively. We see that at the same intensity
of the pulsar emission, the pulse profiles are almost
identical irrespective of the outburst phase.

   It is interesting to trace the evolution of the pulse
profile shape with energy: in soft ($<20$ keV) energy
channels, the profile is double-peaked with a tendency
for the second peak to disappear as the pulsar 
luminosity decreases; as the energy increases, the second
peak also disappears and the profile becomes virtually
single-peaked above $\sim20$ keV, with the width of the
first peak being $\sim0.5$ of the phase, but the width of
the first peak at about 30 keV increases and reaches
$\sim0.75$ of the phase, partially covering the region of
the first peak. A distinctive feature of the average
pulse profile for the source under study is the presence
of several peculiarities in its shape--additional small
peaks (e.g., at phases 0.14 and 0.92 in the 
observations with a luminosity $<7\times10^{37}$ erg s$^{-1}$) and
an asymmetry of the second (with a lower intensity)
peak.

    To qualitatively explain the observed behavior of
the pulse profile (the decrease in the intensity of the
second peak with decreasing luminosity and 
increasing energy), we can suggest a simple, purely 
geometrical picture that is capable of describing the main
observed trends in the pulse profile in general terms:
the rotation axis of the neutron star is inclined with
respect to the axis of its magnetic field in such a way
that the accretion column at one of the poles is seen
over its entire (or almost entire) height when the pole
falls on the observer's line of sight; only the upper
part emitting softer photons is seen in the second
column, while the emission region of hard photons
is screened by the neutron-star surface (hence the
observed decrease in the intensity of the second peak
with increasing energy); as the accretion rate and,
accordingly, the source luminosity decrease, the 
column height decreases, the intensity of the second
peak falls, and we will cease to see it altogether at
some time.

    Naturally, to describe the observed behavior of
the pulse profiles more or less accurately, we must
complicate significantly the picture described above.
Thus, for example, we must know the temperature
distribution along the accretion column, the shape of
the beam function, its dependence on
the object's luminosity, energy band, etc. In addition,
since the emission region is close to the 
neutronstar surface, we must take into account the general
relativity effects, which are also capable of affecting
the pulse shape (see, e.g., Beloborodov 2002).

   The evolution of the pulse profile shape can be
clearly illustrated by the intensity maps constructed
for the following series of observations: 40411-01-02-00, 
40070-01-03-00, 40070-01-05-00, 40051-05-09-00, 
40051-05-12-00, and 40051-05-13-00
(see Table 1 and Fig. 2). The maps were obtained
by folding the pulsar light curve in narrow (about
4-keV-wide) energy channels whose centroid was
displaced from channel to channel by 1 keV. Each
profile was constructed in units relative to the mean
count rate in a given channel. The resulting map
is shown normalized to unity (all intensities were
divided by the maximum value over the entire map).
The figure shows the maps obtained from PCA and
HEXTE data for the first three observations and
only from PCA data for the sessions with a lower
luminosity. The dashed lines indicate the positions
of the fundamental harmonic in the pulsar spectrum
(the positions of the first three harmonics for the
sessions with a high luminosity). This method of
analysis suggested by Tsygankov et al. (2006) when
investigating the pulsar V0332+53 allows one to
detect and trace large-scale variations in the pulse
profile shape with energy and phase. In particular, a
"wavy" behavior of the profile shape was found to be
observed at all luminosities of the pulsar 4U 0115+63
as the energy varies. The effect lies in the fact that
each line of equal intensity of the main peak does
not lie at the same phase at different energies, but
is slightly displaced alternately in one and another
directions. Interestingly, this phase variability of the
main peak exhibits a repeatability in energy and its
period roughly coincides with the distance between
the harmonics of the cyclotron feature in the pulsar
spectrum (Fig. 2). As was shown by Tsygankov et al.
(2006), the cyclotron feature can affect significantly
the pulse profile shape immediately near it. If the
observed "waviness" of the pulse profile evolution
is assumed to be actually related to the peculiarities
of the emission near the cyclotron frequency and its
higher harmonics, then one might expect a "wave
phase" shift depending on the pulsar luminosity
simultaneously with the shift of the cyclotron feature
in the spectrum. However, our analysis did not allow
us to detect the presence or absence of this shift at a
statistically significant level.

{\it The Pulse Fraction}

    Our intensity maps give a qualitative picture of
the pulse profile behavior. To quantitatively describe
the observed variations, we used the dependence of
the pulse fraction, which is defined as $P=(I_{max}-I_{min})/(I_{max}+I_{min})$,
where $I_{min}$ and $I_{max}$ are the
background-corrected count rates at the pulse profile
minimum and maximum, on energy and luminosity.
In all observations, the pulse fraction, on average, 
increases with energy. The typical energy dependence of
the pulse fraction obtained in a wide energy band, 
3-100 keV, from PCA/RXTE and HEXTE/RXTE data
is shown in Fig. 3 (for the brightest pointing 
40070-01-03-00). Figure 4 shows the energy dependences
of the pulse fraction for observations with various
pulsar luminosities derived from PCA data in the 
3-20 keV energy band; the corresponding luminosities
are indicated to the right of each plot. We see that
the pulse fraction decreases significantly (from $\sim65\%$
to $\sim40\%$ at 10 keV) with increasing luminosity. This
result can be explained qualitatively and understood
in terms of the simple model suggested above, which
describes the luminosity dependence of the pulse
profile shape. As the luminosity rises, the geometrical
sizes of the emitting regions increase and, accordingly, 
the pulsations are "smeared". The increase in
pulse fraction with energy can also be explained by
the fact that the emitting regions are more compact
(Fig. 3).

   It should be noted that, as we see from Fig. 3 and 4,
the pulse fraction increases with energy nonlinearly.
The presented dependences exhibit local maxima and
minima whose occurrence periodicity roughly 
coincides with the periodicity of the "wavy" structure of
the pulse profile evolution (see above) and may be of
the same nature.

    To analyze the pulse profile at high energies (above
20 keV) and to investigate the profile evolution with
energy, we used IBIS/INTEGRAL observations with
a considerably longer exposure time and, accordingly,
a better statistical data quality (see Table 2). The
vertical dashed lines in the average intensity map
constructed from these data (Fig. 5) indicate the 
centers of the second, third, and fourth cyclotron line
harmonics in the pulsar spectrum. We see that the
"wavy" structure of the pulse profile evolution with
energy noted above based on the RXTE data is also
present in a more explicit form in the INTEGRAL 
data. Nevertheless, whether it is related to the position
of the cyclotron line in the source spectrum is still an
open question.

{\it Pulse Profile Variability on the Scale of the Pulsation Period}

    All of the pulse profile properties described above
referred to pulses averaged over a long time interval
(of the order of several thousand seconds). However,
some of the physical and geometrical properties of
the emitting regions and their variations can lead to
pulse profile variability on time scales shorter than
one neutron-star rotation period. In particular, such
variability was found in the X-ray pulsar A0535+26.
It was shown for this pulsar that the profile-averaged
variability cannot be explained only by a Poisson 
process and that the variability of one of the peaks in the
profile is higher than the average variability level over
the entire profile (Frontera et al. 1985).

    To investigate the profile variations from pulse to
pulse, we analyzed the pulsar light curves in 
various energy channels and at various luminosities. We
found that the profile is not stable, but varies 
significantly on a time scale of the order of the pulsation
period. As an example, Fig. 6 shows an arbitrary 
segment of the source light curve in the 17--20.3 keV 
energy band for the observing session 40070-01-05-00,
when the source luminosity was $\sim 6.6\times10^{37}$ erg s$^{-1}$.
The dashed line in the same figure indicates the pulse
profile in the same energy band averaged over the
entire observing session. We see that the pulse shape
is not constant, but changes significantly on the scale
of one neutron-star rotation period, while the second
peak in the profile (with a lower intensity) exhibits a
larger variability in both amplitude and shape.

    To quantitatively estimate the pulse profile 
variability, we analyzed the rms deviation of the count
rate in the light curve from the count rate at the
corresponding phase of the average profile obtained
by folding this light curve with the best period. This
quantity (we will call it $RMS$ for short) was calculated as

\begin{equation}\label{4uac2}
RMS=\frac{\sqrt\frac{\sum_{k=1}^N(P_k-<P_k>)^2-N\sigma^2}{N}}{<Flux>},
\end{equation}

where $P_k$ is the background-corrected count rate in a
given bin of the light curve, $<P_k>$ is the count rate in
the bin with the corresponding phase of the averaged
profile, $N$ is the total number of bins in the light
curve equal to the number of periods that fit in the
observation under consideration multiplied by 25 (the
number of bins into which each pulse was broken
down), $\sigma$ is the mean measurement error of the count
rate in the light curve (this term was introduced to
allow for the Poisson noise), and $<Flux>$ is the mean
count rate for the entire observation.

    The rms deviation $RMS$ obtained in this way 
reflects the mean variability of the source pulse profile
in a given energy band at a given luminosity. To
investigate the variability of a particular 
characteristic feature in the profile, i.e., to analyze the phase
dependence of the pulse profile variability, we 
calculated $RMS$ in a similar way, but the summation was
performed only over the bins with the corresponding 
pulse phases. Figure 7 shows the $RMS$ values
obtained in this way divided by the intensity in a
given bin of the average profile in four energy bands
(upper panels, in \%); for comparison, the lower panels
show the average pulse profiles in the corresponding
energy bands. Such dependences were derived for
all observing sessions, but in Fig. 7 we present the
results only for two of them (observations 40070-01-03-00 
and 40070-01-05-00), when the source was
bright enough and a number of characteristic features
were present in the pulse profile.

    We see from the figure that the ratio of $RMS$ to the
intensity in both observations is not constant and 
behaves in a fairly complex way with pulse phase, but we
can distinguish some of its peculiarities (in particular,
for session 40070-01-03-00). In all energy bands, the
first peak (with a higher intensity) is most stable (the
ratio of $RMS$ to the intensity is at a minimum). Its
intensity varies insignificantly; only the peak width
exhibits a small variability, which is reflected in an
increase in $RMS$ in its "wings" (Fig. 7). In contrast,
the second peak is much more variable in both amplitude 
(its variability as a whole is higher than the mean
level by several percent) and shape: the corresponding
segment of the dependence of $RMS$ on pulse phase
(see the upper panels in Fig. 7) manifests itself as a
double- or multiple-peaked structure whose variability 
at maxima is higher than the mean one by 5--10\%.

    The observed pulse profile variability can be attributed 
to various physical processes: nonstationarity of the 
processes in the accretion flow, intensity
variability of some nonpulsating continuum, intrinsic
variability of the pulse profile, etc. Mathematically,
these variability formation mechanisms can be represented as follows:

   (i) Multiplicative addition of the variability -- the
resulting intensity at the current time is defined as
$I_1=\widehat{F}(t) \times P(\varphi)$, where $\widehat{F}(t)$ is a function of only the
time and $P(\varphi)$ is the intensity in the averaged profile
at phase $\varphi$. If only this mechanism were realized, then
the ratio of $RMS$ to the intensity of the average profile
would be constant.

    (ii) Additive addition of the variability, when the intensity 
is specified as $I_2=F(t) + P(\varphi)$. In this case,
as follows from our definition of $RMS$, its ratio to the
intensity will be inversely proportional to the intensity
of the average profile.

    As we see from Fig. 7 (and, to some extent, from
Fig. 6), the observed variability pattern cannot be fully
explained by any of these scenarios, which reflect the
effect of only external factors and do not affect the
variability of the profile itself, in pure form and by their
linear combination. The most generalized case most
likely takes place for the pulsar 4U 0115+63.

    (iii) $I_3=F(t) + \widehat{F}(t) \times (P(\varphi) +
\widetilde{P}(t,\varphi))$, where
$\widetilde{P}(t,\varphi)$ is the component that reflects the processes
related to the variation of the pulse profile itself.

    As regards the physical causes of the intrinsic profile 
variability ($\widetilde{P}(t,\varphi)$), apart from the change in the
shapes and positions of hot spots on the neutron-star
surface (see, e.g., Romanova et al. 2004), the complex
structure of the second peak can result from magnetic
field multipolarity near the neutron-star surface. The
fact that the mean profile variability increases with
energy (Fig. 7) can also serve as circumstantial evidence 
that this variability corresponds to the variability of 
the accretion flow in the immediate vicinity of
the neutron-star surface, where the photon energy is
higher.

    Thus, the method that we used to analyze the
pulse profile shape reveals its "true" variability, which
is "blurred" in the average profile, and its stable component. 
The part of the variability responsible for the
"intrinsic" profile variability carries information about
specific physical processes in X-ray pulsars and must
be taken into account when properly modeling the
formation of the pulse profile for such sources.

\section*{SPECTRAL ANALYSIS}

   To describe the source spectrum, we used a combination 
of models that consisted of a power law and a
high-energy cutoff (the powerlaw*highecut model in
the XSPEC package; White et al. 1983) modified by
absorption lines in the shape of a Lorenz profile,
$exp\left(\frac{-\tau_{cycl}(E/E_{cycl})^2\sigma_{cycl}^2}{(E-E_{cycl})^2+\sigma_{cycl}^2}\right)$,
where $E_{cycl}$ is the energy of the line center, $\tau_{cycl}$ is the
line depth, and $\sigma_{cycl}$ is the line width. The fluorescent
iron line at 6.4 keV was also detected in the spectrum
when fitting the data obtained in observations with a
high emission intensity.

    It should be noted that whether the iron emission 
line is present in the source spectrum is still an
open question. In several papers (Tamura et al. 1992;
Nakajima et al. 2006), it was not required to include
this feature in the model to satisfactorily describe
the experimental data. In contrast, another group of
authors (Nagase et al. 1991; Lutovinov et al. 2000;
Mihara et al. 2004) detected this feature at a statistically 
significant level. This ambiguity results from
the fact that the analytical model used is complex
and somewhat "artificial" and that the cutoff energy
in the source spectrum $E_{cut}$ is close to the line energy.
The final conclusion about the presence of this feature
in the pulsar spectrum is difficult to reach, but its
inclusion in the fitting model reduces significantly the
$\chi^{2}$ value. The line equivalent width for different observations 
changes from $\sim150$ eV in a state with a high
luminosity to $\sim50$ eV in a state with a low luminosity.
This value is compatible with the situation where
the "cold" matter behind the X-ray source (e.g., the
matter flowing over the neutron-star magnetosphere
or the bent edge of the accretion disk) intercepts and
reradiates in the observer's direction about half of the
incident emission (George and Fabian 1991). Based
on this fact, we can interpret the decrease in the
iron line equivalent width with declining luminosity
as a decrease in the area of the reflecting surface.
Much information about the geometry of reflecting
regions and, in particular, the distances to them can
be obtained by analyzing the behavior of the iron line
with pulse phase. However, due to the complex shape
of the spectrum and the relatively low intensity of this
line, observations with a high energy resolution near
6.4 keV are required to detect it.

   Figure 8 shows the source spectrum in a wide
energy band obtained from INTEGRAL data. Note
that the quality of the fit to the experimental data at
energies below 20 keV is not ideal; in particular, this
may stem from the fact that the JEM-X response
matrix is imperfect (for comments, see Filippova et al.
2005). The parameters of the model consisting of a
power law with a high-energy cutoff and four cyclotron 
line harmonics are given in Table. 3.

\begin{center}
{\bf Table 3. Spectral parameters of 4U 0115+63 with INTEGRAL.}

\begin{tabular}{ll}
\hline
Model parameters & Value  \\

\hline

Photon index           & $0.093^{+0.007}_{-0.001}$ \\
$E_{cut}$, keV         & $8.93\pm0.03$ \\
$E_{fold}$, keV        & $9.06^{+0.09}_{-0.02}$ \\
$\tau_{cycl,1}$        & $0.55\pm0.01$ \\
$E_{cycl,1}$, keV      & $11.16^{+0.03}_{-0.02}$ \\
$\sigma_{cycl,1}$, keV & $3.13^{+0.07}_{-0.02}$ \\
$\tau_{cycl,2}$        & $0.97\pm0.01$ \\
$E_{cycl,2}$, keV      & $21.16^{+0.11}_{-0.02}$ \\
$\sigma_{cycl,2}$, keV & $7.55^{+0.15}_{-0.02}$ \\
$\tau_{cycl,3}$        & $0.40\pm0.01$ \\
$E_{cycl,3}$, keV      & $34.55^{+0.01}_{-0.20}$ \\
$\sigma_{cycl,3}$, keV & $4.5^{+0.3}_{-0.1}$ \\
$\tau_{cycl,4}$        & $0.55\pm0.01$ \\
$E_{cycl,4}$, keV      & $44.93^{+0.15}_{-0.27}$ \\
$\sigma_{cycl,4}$, keV & $11.38^{+0.45}_{-0.17}$ \\
$E_{Fe}$, keV          & $6.4$ fixed\\
$\sigma_{Fe}$, keV     & $0.2$ fixed\\
$EW_{Fe}$, eV               & $130\pm10$ \\
$\chi^2$ (d.o.f)       & $0.85 (155)$\\

\hline
\end{tabular}
\end{center}

   The derived model parameters agree with the results 
obtained from RXTE data and earlier by other
authors.

   The characteristic source energy spectrum obtained 
in two observations with different luminosities
and cyclotron energies (observations 40070-01-05-00 
and 40051-05-09-00) from PCA data are shown
in Fig. 9.

{\it Luminosity Dependence of the Cyclotron Line Energy}

    Mihara et al. (1998) and Nakajima et al. (2006)
showed that the position of the cyclotron absorption
line changes with pulsar luminosity from $\sim11$ keV
near the outburst maximum to $\sim16$ keV at the
end of the outburst, when the source luminosity
falls by more than an order of magnitude. If the
observed emission at the end of the outburst is
assumed to originate mainly near the neutron-star
surface, then we can estimate the magnetic field of
1he neutron star, $B=(1+z)\times
E_{cycl}\times10^{12}/11.6\simeq1.4\times10^{12}$ G. 
Nevertheless, although the pulsar properties 
have long been studied, the cyclotron frequency
variation with source luminosity is not completely
clear. Below, we investigate this transition during the
2004 outburst; for comparison, we also analyze the
data for the 1999 outburst.

     Figures 10 and 11 show the time dependences
of the luminosity (open squares) and position of
the fundamental cyclotron frequency harmonic (the
filled squares and triangles for the 1999 and 2004
outbursts, respectively). It is clearly seen from the
figures that the cyclotron energy at high luminosities
($\sim10^{38}$ erg s$^{-1}$) lies within the range $\sim10-11$ keV
and is virtually independent of the pulsar emission 
intensity. As the luminosity decreases to $\sim5\times10^{37}$
erg s$^{-1}$, the cyclotron energy increases sharply
to $\sim14$ keV, with its small rise to $\sim15-16$ keV as the
source luminosity decreases further.

    The cyclotron frequency is plotted against the intrinsic 
luminosity in Fig. 12 based on RXTE (the
designations are the same as those in Figs. 10 and 11)
and INTEGRAL (filled circle) data. We see that the
results obtained with different instruments and during
different outbursts are in good agreement. In both
cases, the cyclotron line energy changes sharply at
the same luminosity ($\sim5\times10^{37}$
erg s$^{-1}$), suggesting
that the observed jump is probably a fundamental
property of the pulsar under study. Searching for and
investigating of such a transition at the outburst rise
phase are of great interest. However, the observations
during both outbursts were begun too late and the
characteristic luminosity at which the jump in cyclotron 
line energy is detected was not observed at
the rise phase.

    Assuming a dipole magnetic field configuration, a
relative change in the fundamental harmonic energy
$\Delta E_{cyc}/E_{cyc}
\sim 60\%$ corresponds to a relative change
in its formation height $\Delta R /R \sim 20\%$, which is equivalent 
to a change in the radius by $\sim2$ km (taking the
neutron-star radius to be 10 km).

   By studying the spectral characteristics of the
emission at a certain source luminosity, we can obtain
constraints on the height of the accretion column
segment where this emission is generated effectively.
To a first approximation, the electron energy (Landau)
levels at a given magnetic field strength may be
assumed to be arranged according to the harmonic
law (1:2:3...). The model by Basko and Sunyaev
(1976) predicts that harder radiation in the accretion
column emerges from regions closer to the neutron-star 
surface, where the magnetic field strength is
higher. Thus, the deviations of the energies of the
fundamental and higher harmonics in the spectrum
of 4U 0115+63 from a linear law can be used to
compare the effective sizes of the emitting regions at
$\sim$ 11, 22, 33, and 44 keV, respectively. The energies of
the higher harmonics should lie above the
harmonic law. Figure 13 illustrates the picture described 
above for the observations with luminosities
of $7.4\times10^{37}$ erg s$^{-1}$ from INTEGRAL data (filled
squares and thick lines) and $11.0\times10^{37}$ erg s$^{-1}$ from
RXTE data (open circles and thin lines). Based on the
energy of the fundamental cyclotron line harmonic,
we determined the harmonic law (dotted lines) and
analyzed the deviations of the centroids of the higher
harmonics from it. Since the formal statistical errors
in the line centroid are much smaller than the energy
resolution of the instruments used, we introduced a
systematic uncertainty at the level of a characteristic
scatter of values ($\sim1$ keV). For this purpose, 0.5 keV
was added quadratically to the statistical error.

    To determine the possible scatter of heights in the
accretion column by the method described above, we
took the observed deviations of the centroid energies
for the higher cyclotron line harmonics from the harmonic 
law indicated by the dashed line. Given the
possible statistical and systematic errors, the maximum 
possible deviation from the harmonic law that
is formally compatible with the observational data is
indicated in the figure by the solid lines for each observation. 
This maximum deviation specifies the most
conservative constraint on the deviation from the harmonic 
law. As a result, the observed deviation from
the harmonic law at a luminosity of $7.4\times10^{37}$ erg s$^{-1}$
(INTEGRAL data) was found to be $\Delta
E_{cyc}/E_{cyc} \sim 1\% \pm 3\%$ at the energy of the highest observable
(fourth) cyclotron line harmonic ($\sim45$ keV). Note
that, given the measurement errors, this result is
completely compatible with a purely harmonic dependence 
of the line centroid positions. In terms of the
effective heights of the emitting regions (at energies $\sim$
11, 22, 33, and 44 keV), these deviations correspond 
to $\Delta R /R \sim 0.3\% \pm 1\%$. The most conservative estimate 
yields $\Delta E_{cyc}/E_{cyc} < 9\%$ and $\Delta R
< 300$ m. Consequently, the effective sizes of the emitting 
regions in this state do not differ by more than
several hundred meters and this difference is most
likely even smaller.

   For the other observation (at a higher luminosity, 
$11.0\times10^{37}$ erg s$^{-1}$, RXTE observations),
the observed deviation from the harmonic law is
$\Delta
E_{cyc}/E_{cyc} \sim 16\% \pm 3\%$ at the energy of the highest 
observable (third) cyclotron line harmonic ($\sim36$ keV),
while the corresponding most conservative limit is
$\Delta E_{cyc}/E_{cyc} < 26\%$ ($\Delta R <900$ m). We see from the
figure that in this case the higher harmonics actually
have energies above the harmonic law, as might be
expected at an appreciable linear size of the column
and a dependence of the hardness of the emergent
photons on the distance to the stellar surface.

    For the first observation (at a luminosity $7.4\times10^{37}$ erg s$^{-1}$
appreciably lower than the maximum one), 
the small deviations from the harmonic law
can be interpreted as resulting from the fact that the
emitting region is compact or that the hardness of
the emergent photons does not depend on the distance 
to the stellar surface. Interestingly, the constraints 
on the scatter $\Delta R$ are considerably smaller
than the estimate obtained by assuming that the
abrupt change in cyclotron frequency (at a luminosity
of $\sim5\times10^{37}$ erg s$^{-1}$) is related to a change in the
height of the shock wave in the accretion column.

   At fairly high harmonic energies, the relativistic
corrections to the harmonic law may prove to be
significant (Harding and Daugherty 1991). The correction 
is $\frac{1}{2}\frac{E}{m_ec^2}$ in order of magnitude, i.e., does
not exceed 4\% for the maximum cyclotron line energy
of $\sim$45 keV observed for the source 4U 0115+63.
Therefore, the relativistic corrections have virtually no
effect on the conclusions reached above.

{\it Constancy of the Cutoff energy ($E_{cut}$) in the Source Spectrum}

    Another distinctive feature of the pulsar
4U 0115+63 is the stability of its spectral shape
during an outburst. In particular, the cutoff energy
in the spectrum ($E_{cut}$) does not correlate with the
cyclotron frequency and, being fixed at 8.9 keV, it
distorts only slightly the other parameters of the fit
and does not deteriorate the $\chi^{2}$ value. As an example,
let us consider two RXTE observations, 40051-05-05-00 
and 40051-05-09-00, when the energies of the
fundamental cyclotron line harmonic were $\sim10.7$ and
$\sim14.9$ keV, respectively. Thus, if we fit the chosen
sessions by a model where the cutoff energy $E_{cut}$ is
a free parameter, then the total $\chi^{2}(N)$ value (with $N$
degrees of freedom) will be $16.58(36)$ and $26.08(36)$,
respectively. If, however, $E_{cut}$ is fixed at $8.9$ keV, then
the $\chi^{2}(N)$ values will be $17.16(37)$ and $26.30(37)$,
respectively. This result is interesting in that it allows
us to test the assumption that the cutoff energy in
the spectrum depends on the cyclotron energy. A
correlation between these parameters derived from
the average spectra of various sources was found in
a number of papers (see, e.g., Makishima et al. 1999;
Orlandini and Dal Fiume 2001; Coburn et al. 2002).
Because of the wide dynamic range of the relative
change in cyclotron energy for the source under study
($\sim50\%$), we can check whether this correlation exists
not for different sources, but for the same pulsar.
Using the same set of observations (40051-05-05-00 
and 40051-05-09-00) and assuming that the
cutoff energy in the spectrum in a brighter state
(40051-05-05-00) is $\sim8.9$ keV, $E_{cut}$ in the fitting
model of the second spectrum was specified to be
proportional to the cyclotron energy. As a result, the
quality of the fit proved to be unacceptable, which,
in particular, manifests itself in a sharp increase in
$\chi^{2}(N)$ ($57.35(37)$).

     Since $E_{cut}$ and the remaining parameters of the
fitting model remain essentially constant, we can assume 
that the cutoff energy depends not on the instantaneous 
cyclotron energy (i.e., on the specific
configuration of the accretion regions at the poles),
but rather reflects the fundamental properties of the
neutron star. For example, when analyzing the observations 
of the X-ray pulsar Her X-1, Gruber et al.
(2001) found the cyclotron energy in the source spectrum 
to increase. This was interpreted as a change in
the magnetic field strength of the neutron star. In this
case, a direct proportionality between the cutoff energy 
($E_{cut}$) and the cyclotron frequency was observed.

\section*{CONCLUSIONS}

    We investigated the accreting X-ray pulsar
4U 0115+63 using the RXTE and INTEGRAL data
obtained during intense outbursts in 1999 and 2004.
Below, we briefly summarize the most interesting and
important results.

    -- We analyzed the dependence of the pulse profile 
on the source luminosity; we showed that the
intensity of the second peak in the profile decreases
with declining source luminosity as well as with increasing 
energy and the peak disappears almost completely 
at energies above $\sim20$ keV. We suggested
a model describing qualitatively this behavior of the
profile in which the lower part of one of the accretion 
columns (emitting harder photons) is partially
screened from the observer by the neutron star surface; 
as the source luminosity declines, the column
height decreases and an increasingly large part of it
is screened from the observer.

    -- The pulse fraction was shown to increase both
with decreasing intrinsic source luminosity and with
increasing energy. This suggests that the hard X-ray
emitting regions are more compact, which is compatible 
with the model described above. On the scale
of the pulsar period, we revealed a component in the
variability of the pulse profile shape that is not described 
by the accretion flow variability, but is related
to the intrinsic variability of the pulse profile.

    -- We detected a cyclotron absorption line and its
three higher harmonics in the pulsar spectrum. We
analyzed the dependence of the cyclotron absorption
line energy on the pulsar luminosity and showed that
this dependence is nonlinear and that the line energy
increases abruptly when the source reaches a luminosity 
of $\sim5\times10^{37}$ erg s$^{-1}$. The possible scatter of
heights at which emission with different energies is
produced in the accretion column was estimated from
the deviations of the centroids of the higher cyclotron
line harmonics from an equidistant distribution.

    -- We showed that the cutoff energy in the spectrum 
is be essentially constant during an outburst and
does not correlate with the cyclotron energy.

\section*{ACKNOWLEDGMENTS}

   We thank M.G. Revnivtsev for help with the
RXTE data analysis and for a discussion of the results
obtained. This work was supported by the Ministry of
Industry and Science (Presidential grant RF NSh-1100.2006.2), 
the "Origin and Evolution of Stars and
Galaxies" Program of the Presidium of the Russian
Academy of Sciences, and the Russian Foundation 
for Basic Research (projects no. 05-02-17465 and 07-02-01051).
A.A. Lutovinov thanks the Russian Science Support
Foundation. We used data from the High Energy
Astrophysics Science Archive Research Center Online 
Service provided by the NASA/Goddard Space
Flight Center, the INTEGRAL Science Data Center
(Versoix, Switzerland), and the Russian INTEGRAL
Science Data Center (Moscow, Russia).

\pagebreak

\pagebreak

\clearpage

\newpage

\begin{figure*}[t]
\centerline{\includegraphics[width=16cm,bb=70 75 565 775]{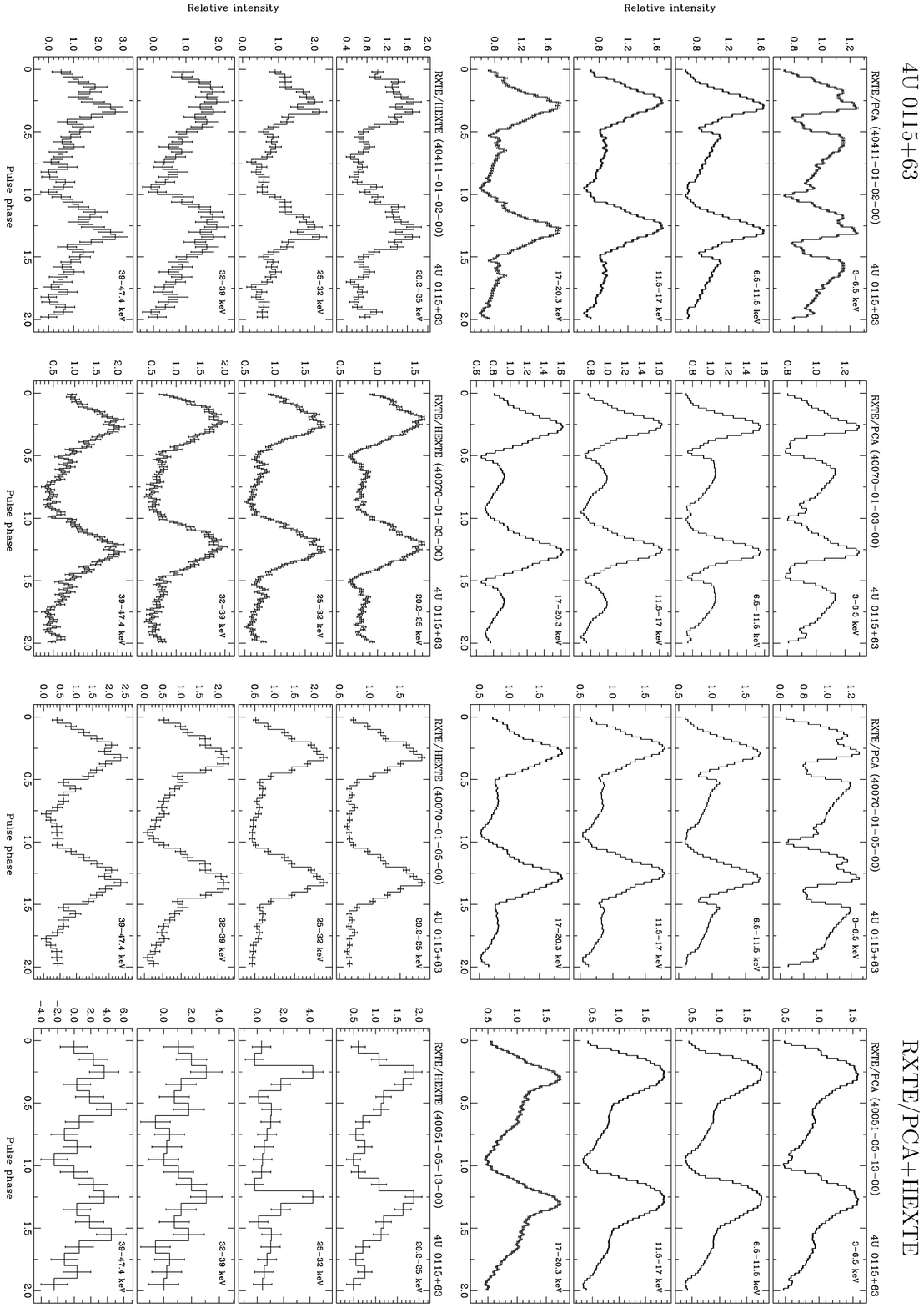}}

\vfill

\caption{Pulse profile shape as a function of the source luminosity and energy band. The columns are arranged from the outburst
onset from left to right and correspond to luminosities $\sim 7.3\times10^{37}$, $\sim 14.6\times10^{37}$,
$\sim 6.6\times10^{37}$, и $\sim 1.5\times10^{37}$ erg s$^{-1}$. The
background was subtracted.}\label{pprof}
\end{figure*}

\newpage
\begin{figure*}[t]
\centerline{\includegraphics[width=16cm,bb=55 60 565 775]{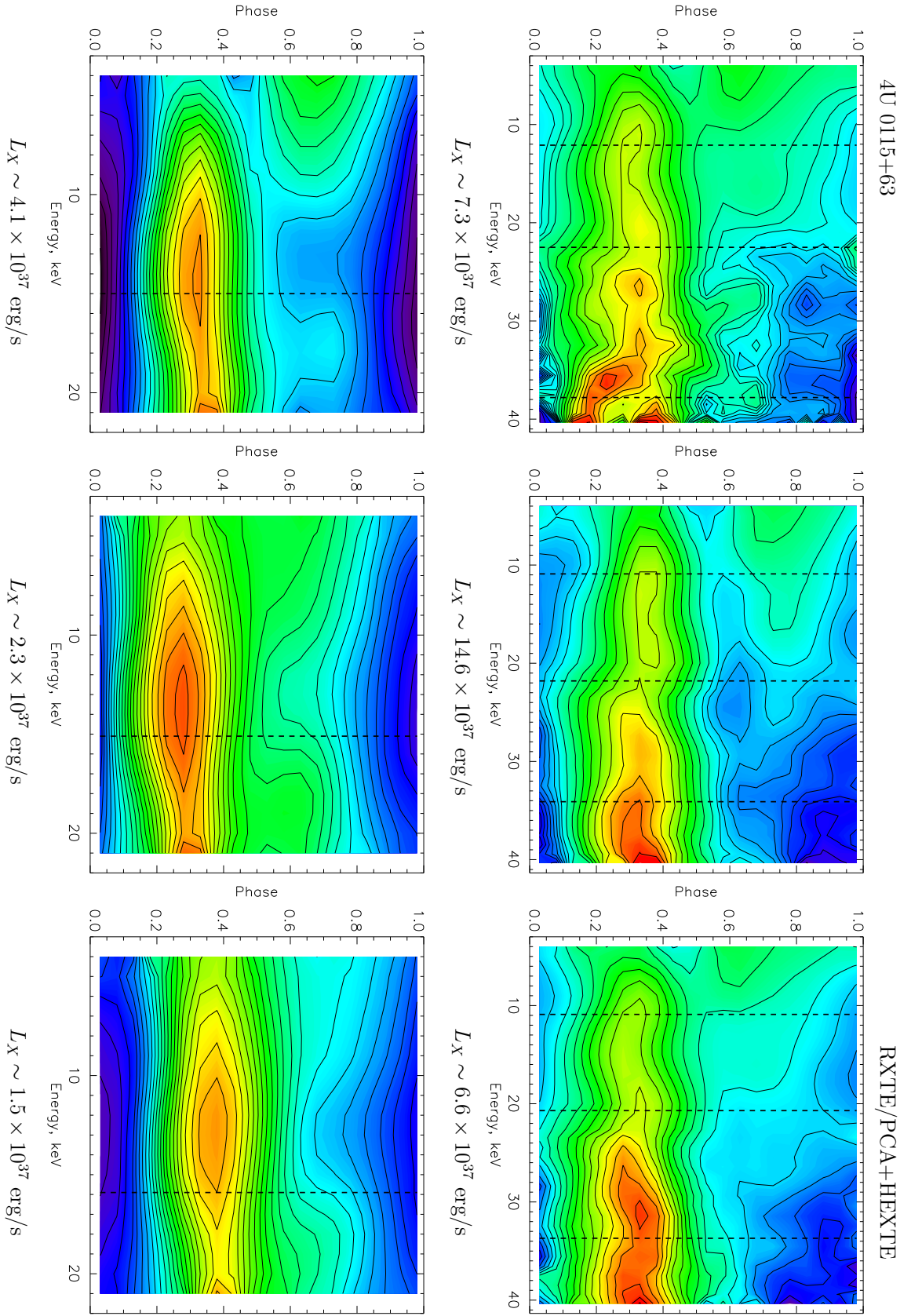}}

\vfill

\caption{Background-corrected intensity maps of the pulse profile for the X-ray pulsar 4U 0115+63 from PCA/RXTE and
HEXTE/RXTE data as a function of the energy band and source bolometric luminosity (see the text). The dashed lines
indicate the positions of the cyclotron frequency harmonics in the source spectrum; the corresponding source luminosities
are given under the maps.}\label{2dprof}
\end{figure*}
\newpage
\begin{figure*}[t]
\centerline{\includegraphics[width=14cm,bb=30 280 510 690]{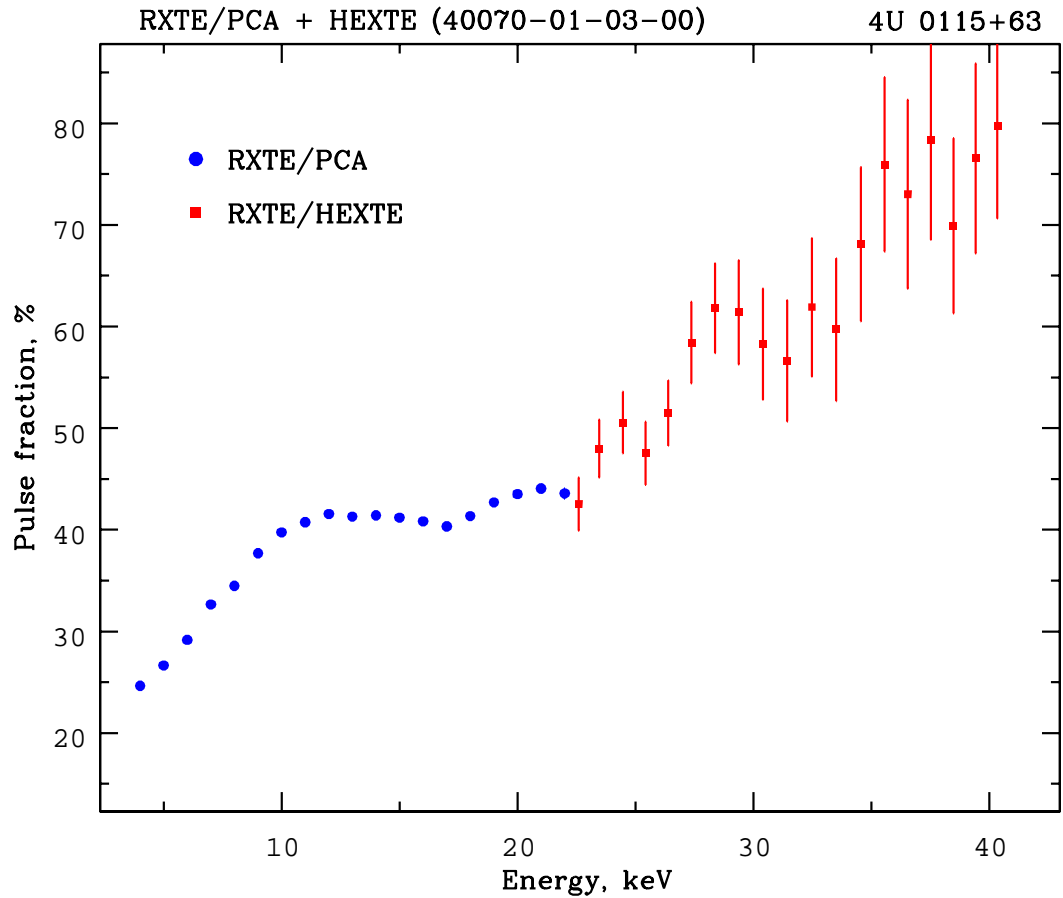}}

\vfill

\caption{Pulse fraction versus energy from the RXTE data
obtained during the 1999 outburst, when the pulsar luminosity 
was at a maximum (observation 40070-01-03-00).}\label{ppfr}
\end{figure*}
\newpage
\begin{figure*}[t]
\centerline{\includegraphics[width=14cm,bb=30 280 510 690]{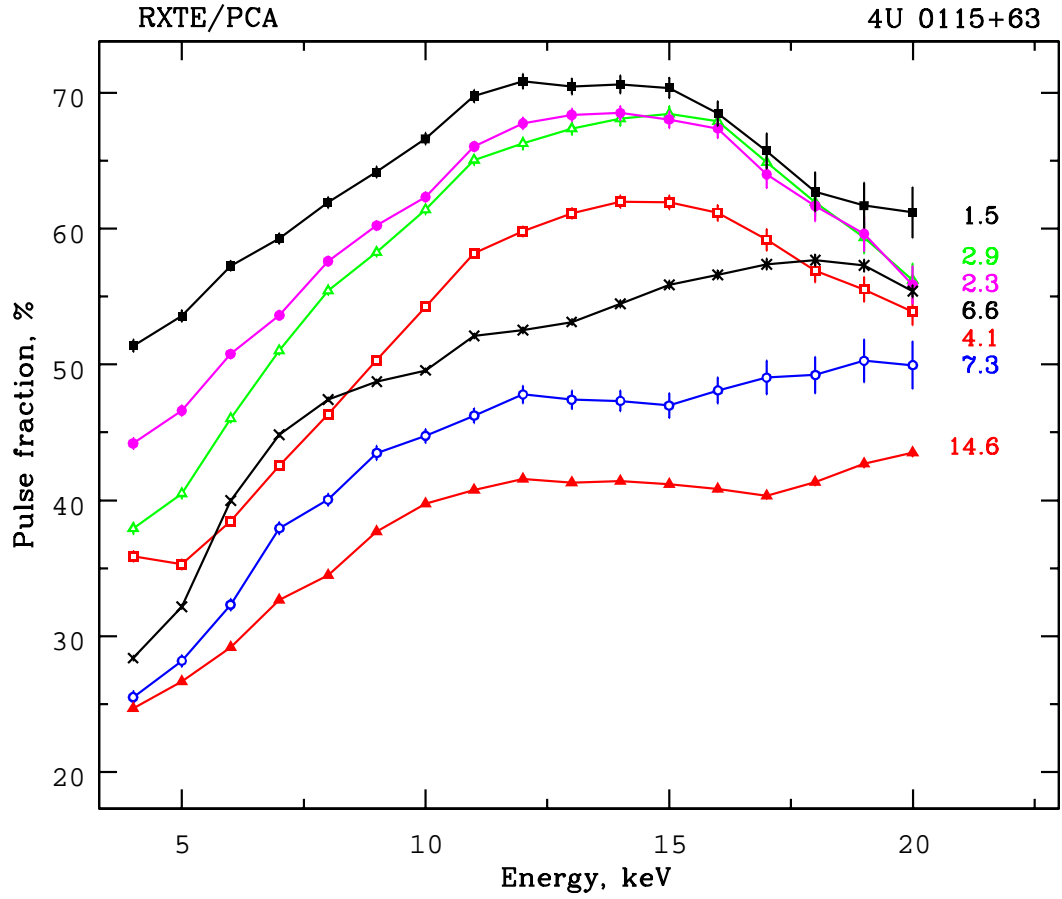}}

\vfill

\caption{Pulse fraction versus energy for various bolometric luminosities of the pulsar (shown to the right of the corresponding
curves in units of $10^{37}$ erg s$^{-1}$) from PCA/RXTE data.}\label{ppfrall}
\end{figure*}
\clearpage
\begin{figure*}[t]
\centerline{\includegraphics[width=14cm,bb=45 310 515 725]{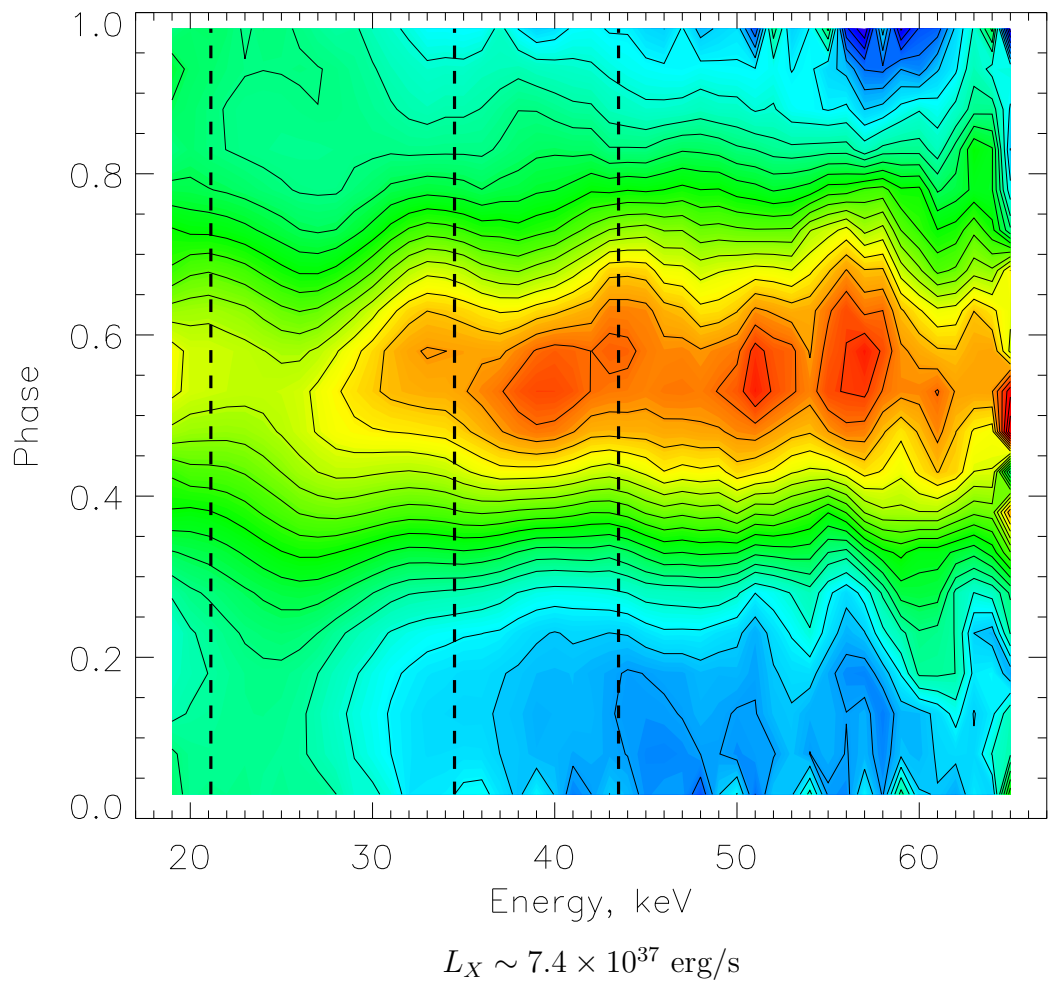}}

\vfill

\centerline{~~~~~~~~~~~~$L_X\sim7.4\times10^{37}$ erg/s}

\centerline{}
\centerline{}

\caption{Background-corrected intensity map of the pulse profile for the X-ray pulsar 4U 0115+63 from IBIS/INTEGRAL
data. The source luminosity in this observation is $\sim 7.4\times10^{37}$ erg s$^{-1}$. The dashed lines indicate the positions of the second,
third, and fourth cyclotron frequency harmonics in the source spectrum.}\label{contint}
\end{figure*}
\newpage
\begin{figure*}[t]
\centerline{\includegraphics[width=12cm,bb=35 290 600 705,clip]{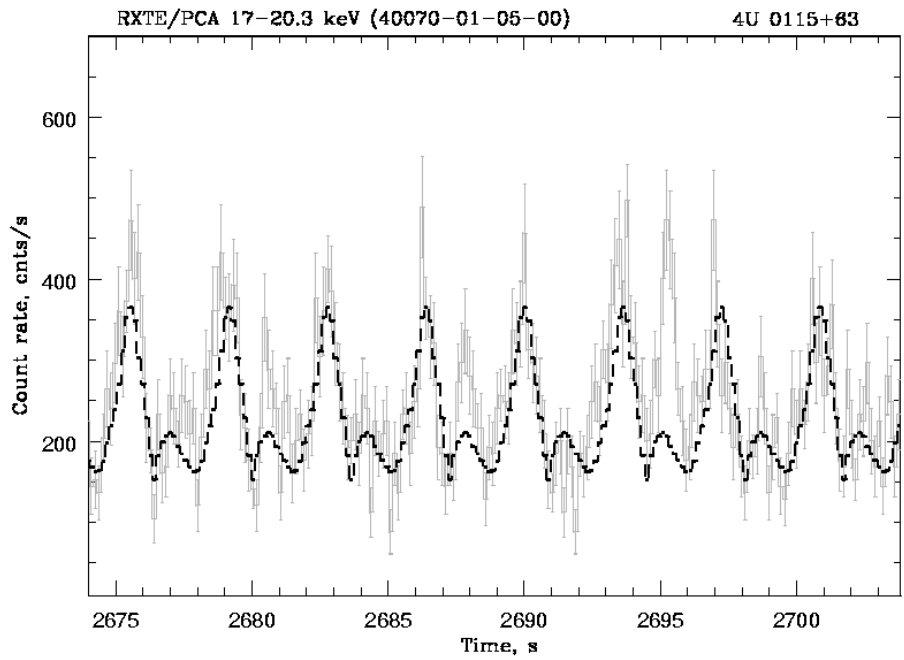}}

\vfill

\caption{Pulsar 17­20.3 keV light curve at a luminosity of $\sim 6.6\times10^{37}$ erg s$^{-1}$ (observation 40070-01-05-00). The dashed line
indicates the average pulse profile shape in this energy band for the entire observing session. The background was subtracted.}\label{lcur}
\end{figure*}
\newpage
\begin{figure*}[t]
\centerline{\includegraphics[width=10cm,bb=180 160 540 720,clip]{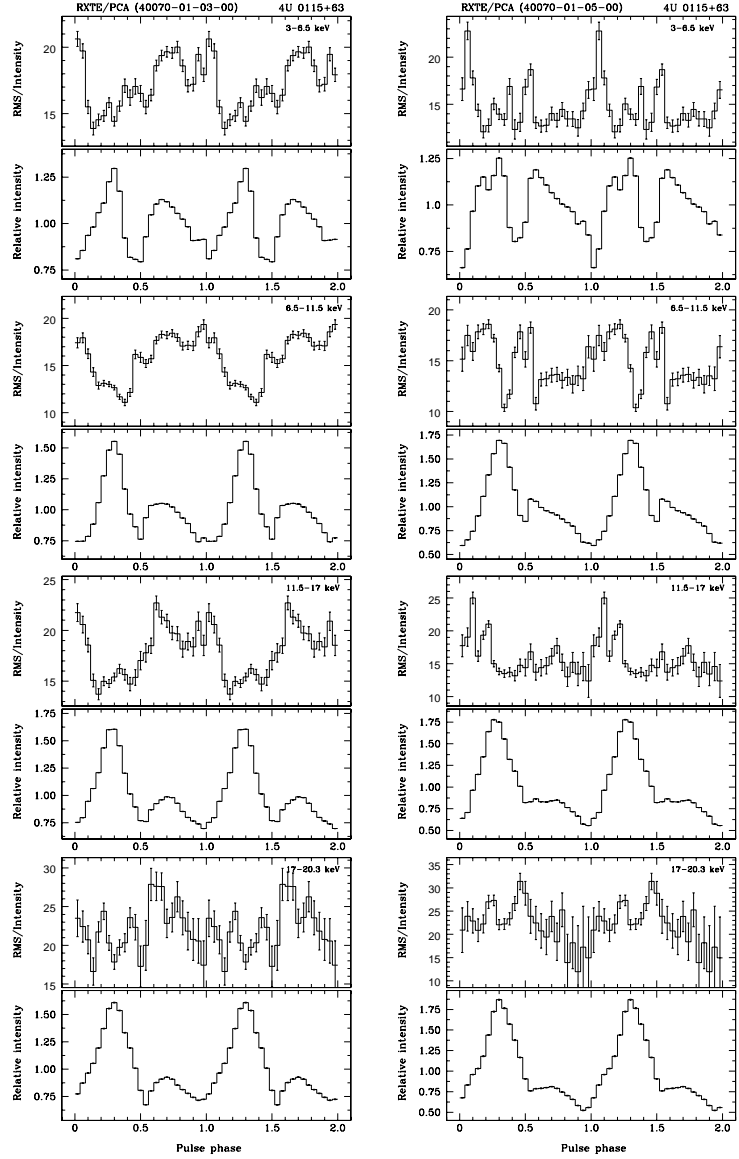}}

\vfill

\caption{Rms deviation ($RMS$ , see the text) divided by the count rate in the average pulse profile versus pulse phase (upper
panels) for two source luminosities ($\sim 14.6\times10^{37}$ and $\sim 6.6\times10^{37}$ erg s$^{-1}$) in various energy bands. For comparison, the
lower panels show the average pulse profiles in the corresponding energy bands for the observations under consideration.}\label{rms}
\end{figure*}
\clearpage

\begin{figure*}[t]
\centerline{\includegraphics[width=14cm,bb=95 240 515 690]{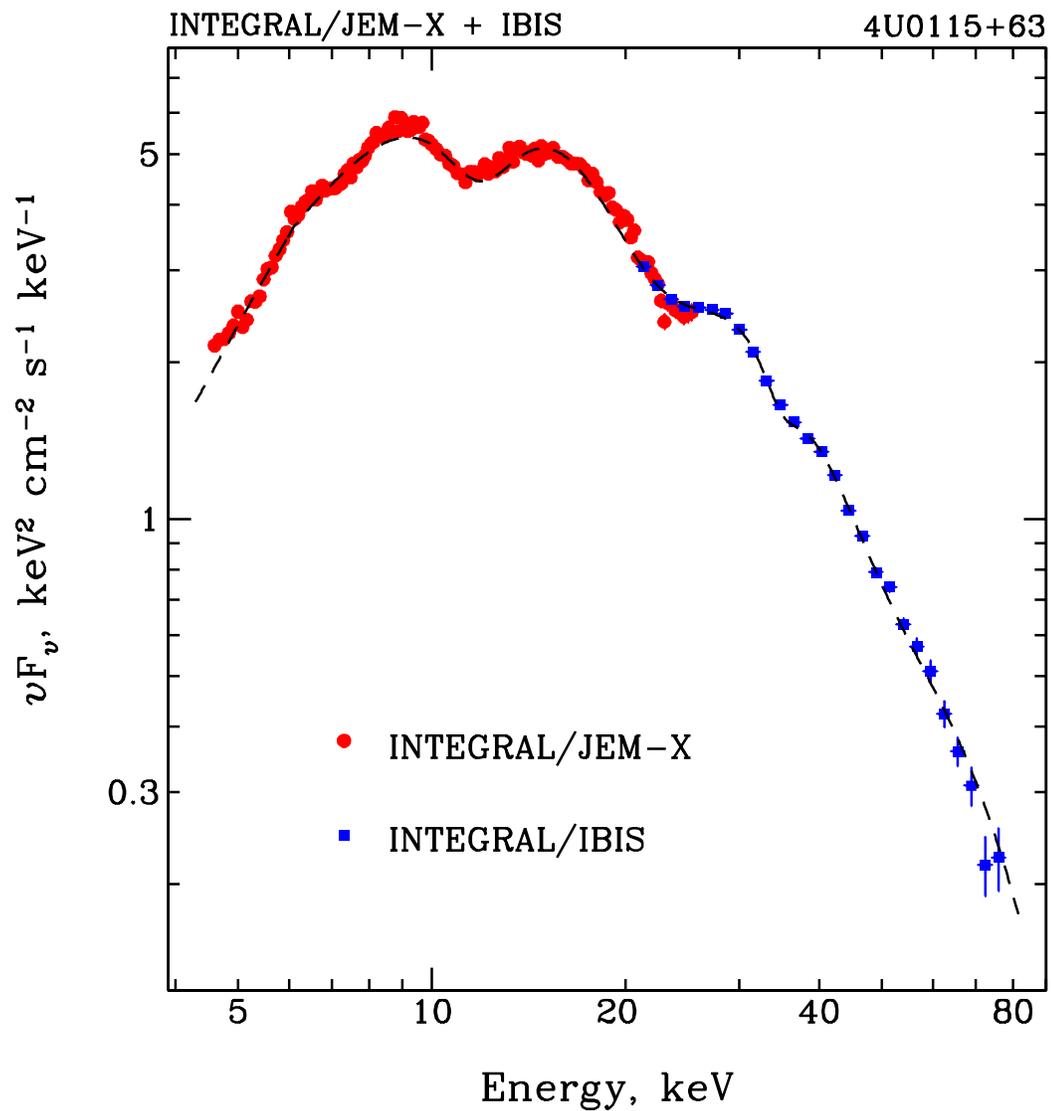}}

\vfill

\caption{Pulsar energy spectrum from INTEGRAL data
in a wide energy band. The model (dashed line) does not
include the component related to the iron emission line
(see the text).}\label{intspec}
\end{figure*}
\newpage
\begin{figure*}[t]
\centerline{\includegraphics[width=14cm,bb=95 240 515 690]{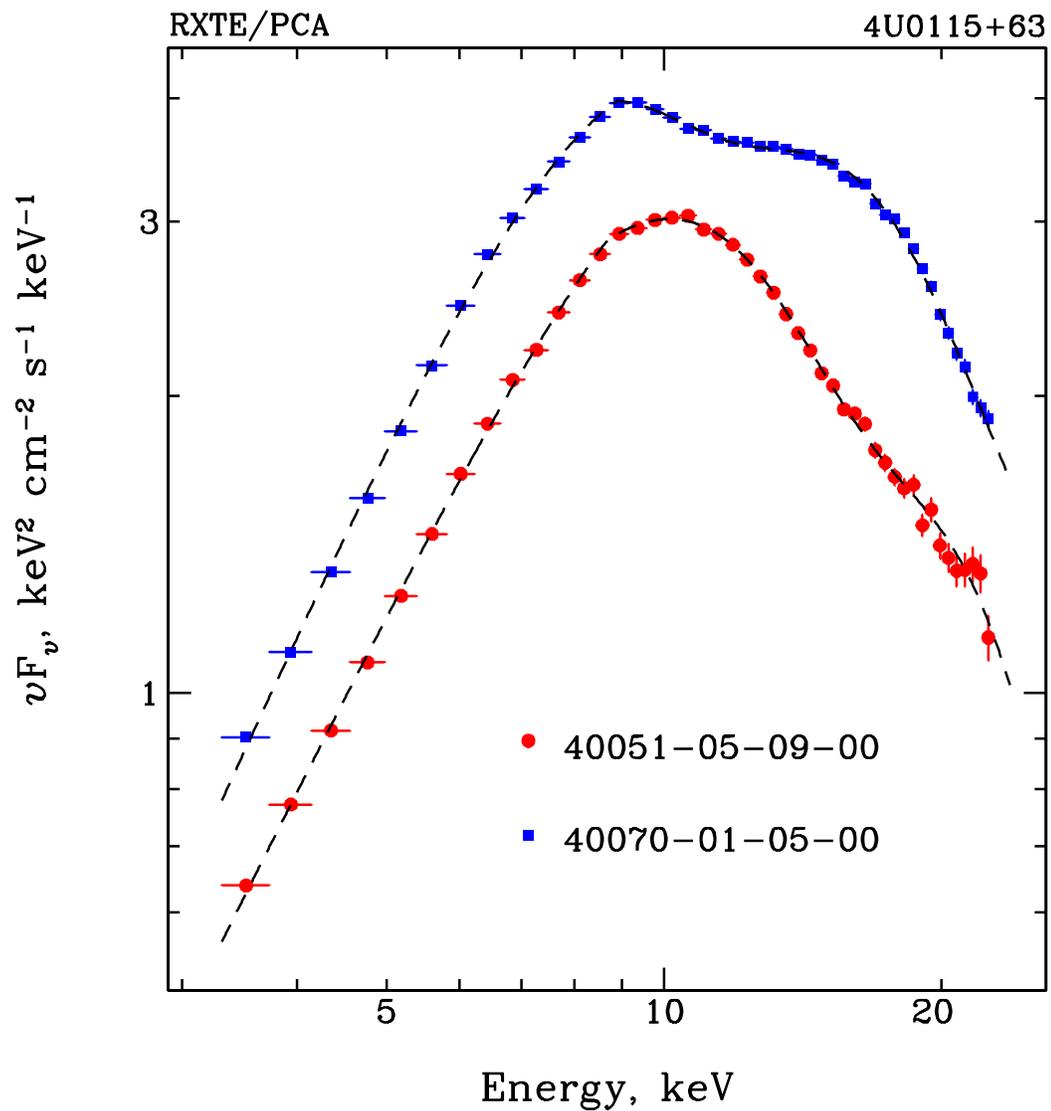}}

\vfill

\caption{Characteristic energy spectra of the source
in states with different luminosities ($\sim 6.6\times10^{37}$ and
$\sim 4.1\times10^{37}$ erg s$^{-1}$) from PCA/RXTE data. The fitting
model (dashed lines) does not include the component
related to the iron emission line.}\label{pcaspec}
\end{figure*}
\newpage

\begin{figure*}[t]
\centerline{\includegraphics[width=14cm,bb=25 280 490 690]{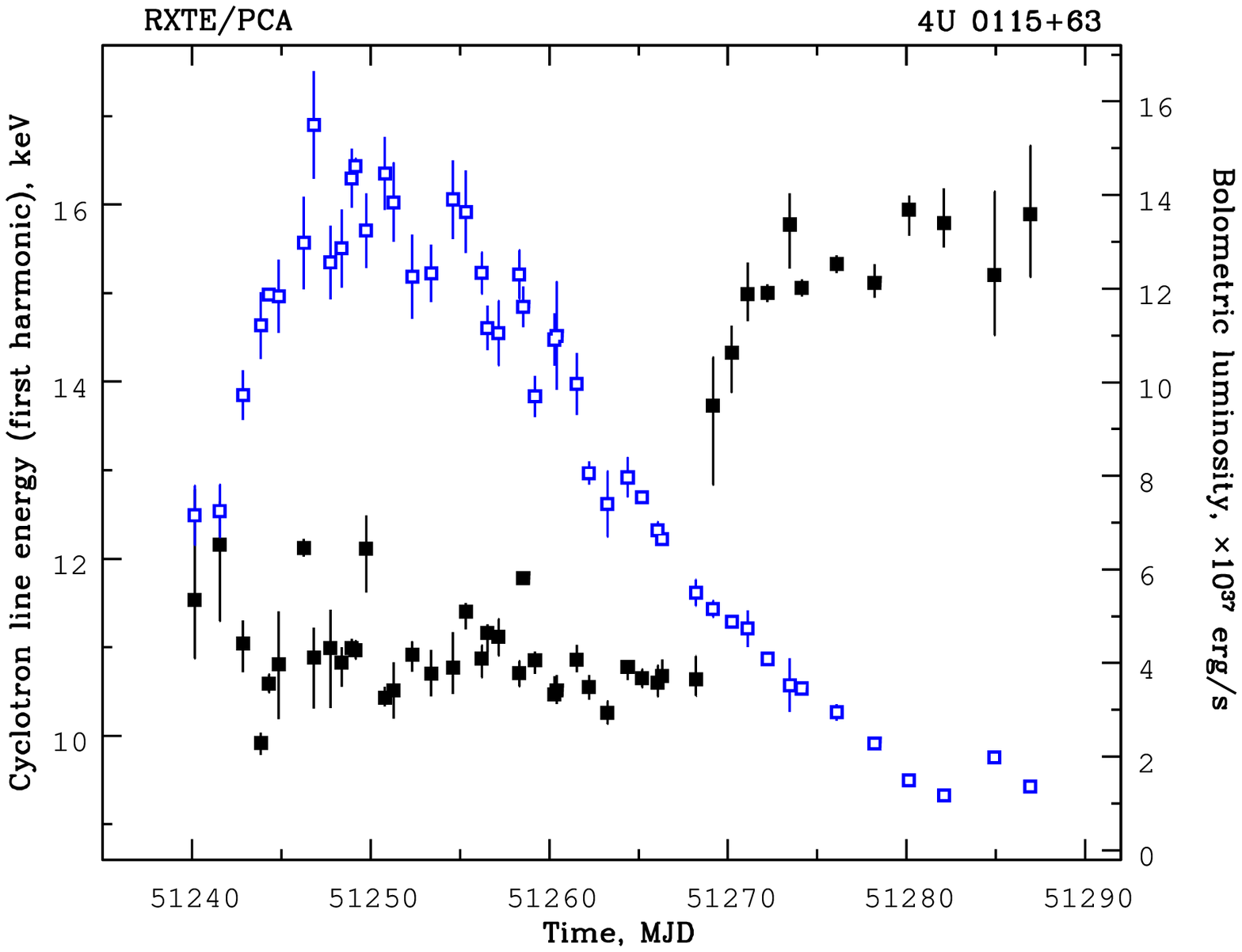}}

\vfill

\caption{Position of the fundamental harmonic of the cyclotron absorption line in the pulsar spectrum (filled squares) and
$3-100$ keV luminosity (open squares) as a function of time from the RXTE data in 1999.}\label{ect}
\end{figure*}
\newpage
\begin{figure*}[t]
\centerline{\includegraphics[width=14cm,bb=25 280 490 690]{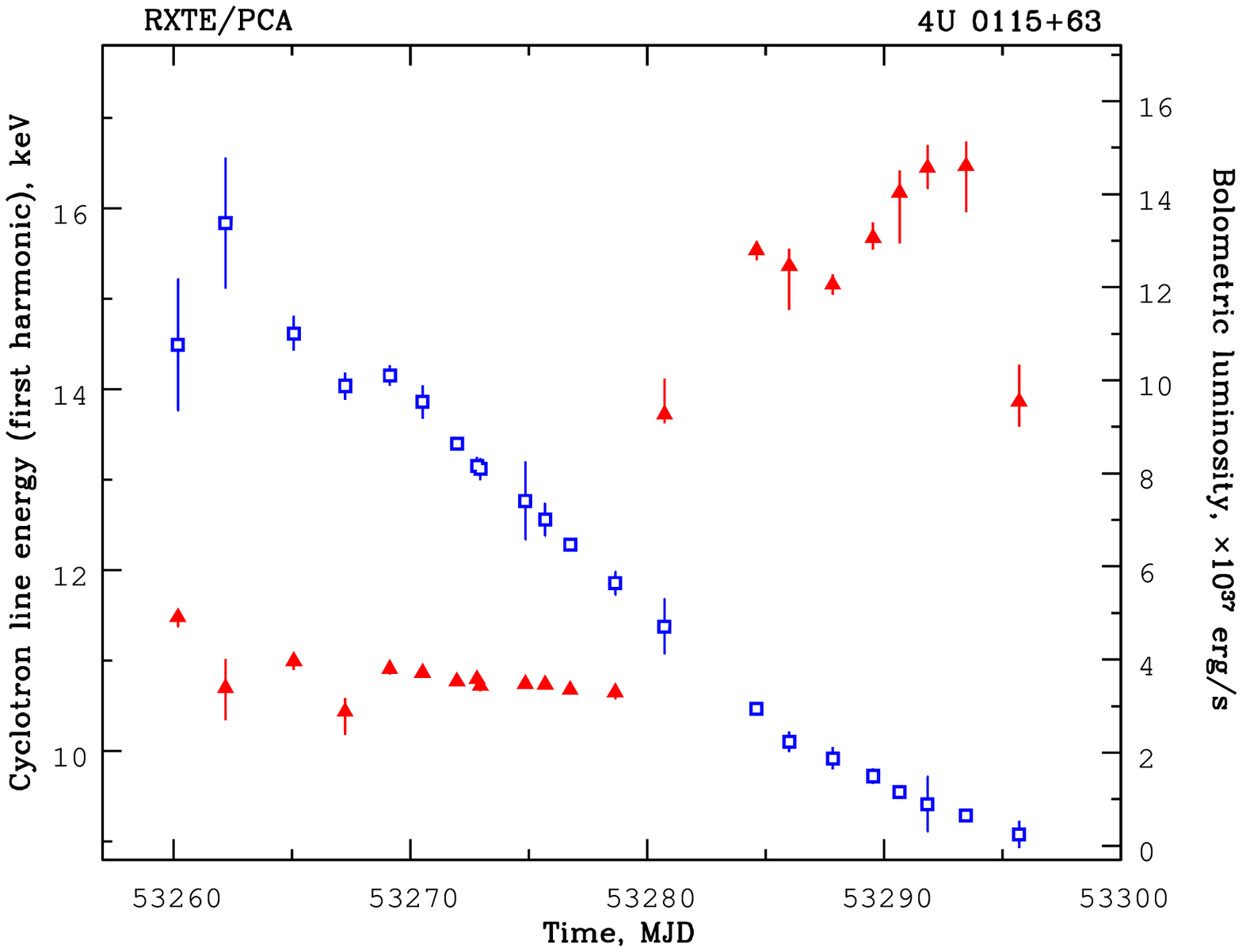}}

\vfill

\caption{Position of the fundamental harmonic of the cyclotron absorption line in the pulsar spectrum (filled triangles) and
$3-100$ keV luminosity (open squares) as a function of time from the RXTE data in 2004.}\label{ect04}
\end{figure*}
\newpage
\begin{figure*}[t]
\centerline{\includegraphics[width=14cm,bb=25 280 515 690]{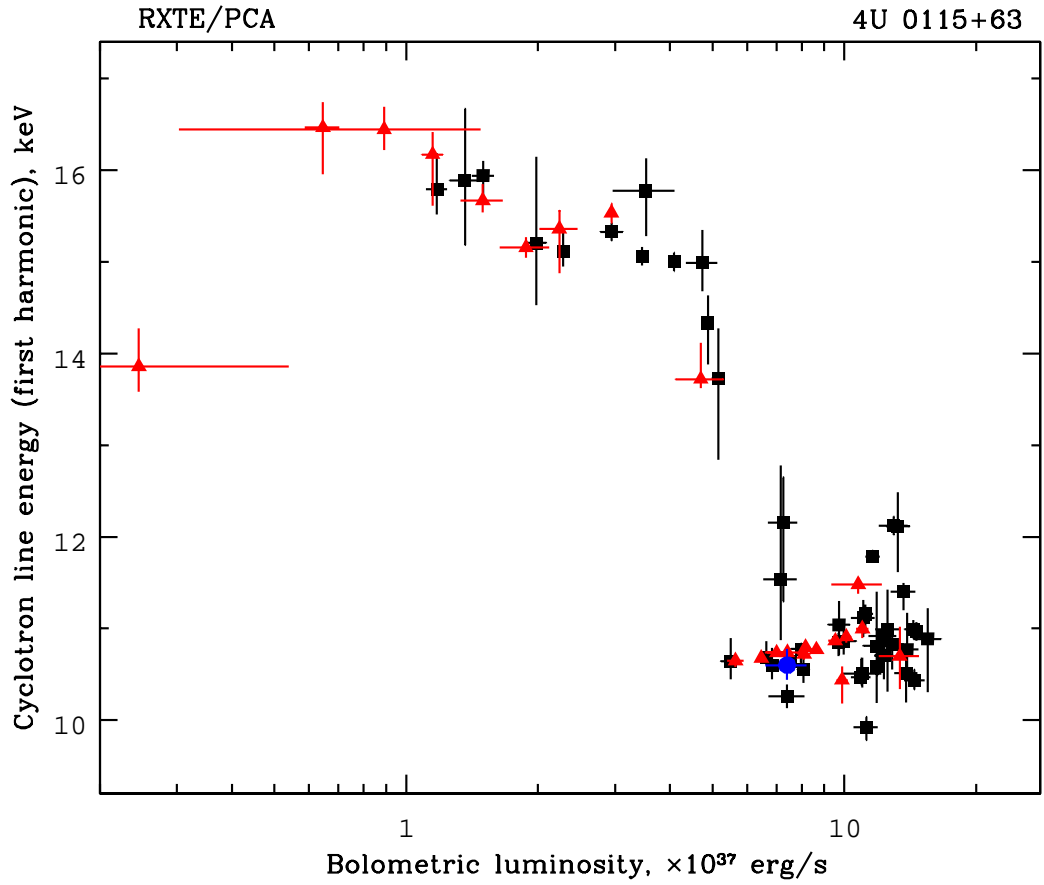}}

\vfill

\caption{Position of the fundamental harmonic of the cyclotron absorption line in the pulsar spectrum as a function of intrinsic
luminosity. The squares, triangles, and circle indicate, respectively, the values during the 1999 outburst from RXTE data,
during the 2004 outburst from RXTE data, and during the 2004 outburst from INTEGRAL data.}\label{ecl}
\end{figure*}
\newpage
\begin{figure*}[t]
\centerline{\includegraphics[width=14cm,bb=25 280 515 690]{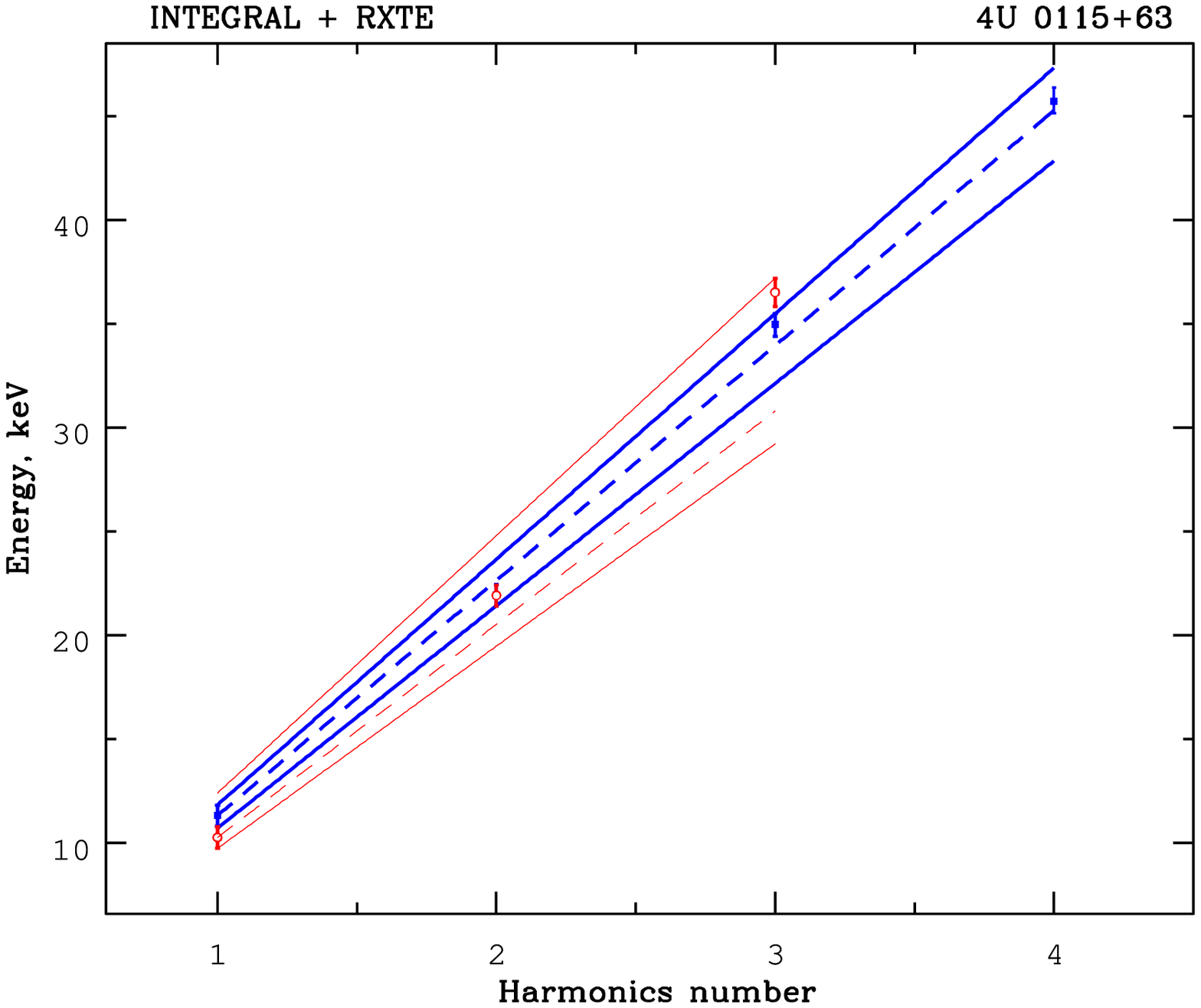}}

\vfill

\caption{Positions of the cyclotron line harmonics in the pulsar spectrum as a function of their numbers for two observations
with different luminosities: $7.4\times10^{37}$ erg s$^{-1}$ from INTEGRAL data (filled squares and thick lines) and $11.0\times10^{37}$ erg s$^{-1}$
from RXTE data (open circles and thin lines). The dashed lines indicate the harmonic arrangement of harmonics; the solid lines
indicate the maximum deviations from it.}\label{harm}
\end{figure*}


\begin{references}

\reference Basko M.M., Sunyaev R.A.,
\mnras\ {\bf 175}, 395 (1976)

\reference Beloborodov A.M.,
\apj\ {\bf 566}, L85 (2002)

\reference Bradt H.V., Rothschild R.E., Swank J.H.,
\aaps\ {\bf 97}, 355 (1993)

\reference Coburn W., Heindl W., Rothschild R. \etal,
\dd\ \apj\, {\bf 580}, 394 (2002)

\reference Cominsky L., Clark G.W., Li F. \etal,  \nat\ {\bf 273}, 367 (1978)

\reference Filippova E., Tsygankov S., Lutovinov A., Sunyaev R.,
Astron. Lett., {\bf  31}, 729 (2005)

\reference Forman W., Jones C., Cominsky L. \etal,
\apjs\, {\bf 38}, 357 (1978)

\reference Frontera F., dal Fiume D., Morelli E., Spada G.,
\apj\, {\bf 298}, 585 (1985)

\reference Gruber D. E., Heindl W. A., Rothschild R. E. \etal,
\apj\, {\bf 562}, 499 (2001)

\reference Giacconi R., Murray S., Gursky H. \etal,
\apj\, {\bf 178}, 281 (1972)

\reference George I.M., Fabian A.C.,
\mnras\ {\bf 249}, 352 (1991)

\reference Harding A.K., Daugherty J.K.,
\apj\, {\bf 374}, 687 (1991)

\reference Heindl W.A., Coburn W., Gruber D.E. \etal,
\apj\, {\bf 521}, L49 (1999)

\reference Hutchings J.B., Crampton D.,
\apj\, {\bf 247}, 222 (1981)

\reference Kholopov P.N., Samus N.N., Kukarkina N.P. \etal,
Information Bulletin on Variable Stars, 2042, 1 (1981)

\reference Lebrun F., Leray J. P., Lavocat P. \etal,  \aap\ {\bf 411}, L141 (2003)

\reference Lund N., Brandt S., Budtz-Joergesen C. \etal,  \aap\ {\bf 411}, L231 (2003)

\reference Lutovinov A. A., Grebenev S. A., Syunyaev R. A., Pavlinskii M. N.,
Astron. Lett., {\bf 20}, 538 (1994)

\reference Lutovinov A. A., Grebenev S. A., Sunyaev R. A.,
Astron. Lett., {\bf 26}, 1 (2000)

\reference Lutovinov A., C.Budtz-Jorgensen, M.Turler \etal,
The Astronomers Telegram {\bf 326} (2004)

\reference Makishima K., Mihara T., Nagase F., Tanaka Y.,
\apj\, {\bf 525}, 978 (1999)

\reference Mihara T., Makishima K., Nagase F.,
\asr\, {\bf 22}, 987 (1998)

\reference Mihara T., Makishima K., Nagase F.,
\apj\, {\bf 610}, 390 (2004)

\reference Mineo T., Ferrigno C., Foschini L.  \etal,  \aap\ {\bf 450}, 617 (2006)

\reference Nagase F., Dotani T., Tanaka Y. \etal,
\apj\, {\bf 375}, L49 (1991)

\reference Nakajima M., Mihara T., Makishima K., Niko H.,
\apj\, {\bf 646}, 1125 (2006)

\reference Negueruela I., Okazaki A.T.,  \aap\ {\bf 369}, 108 (2001)

\reference Orlandini M., Fiume D.Dal,
AIP Conference Proceedings\, {\bf 599}, 283 (2001)

\reference Rappaport S., Clark G.W., Cominsky L. \etal,
\apj\, {\bf 224}, L1 (1978)

\reference Revnivtsev M., Sunyaev R., Varshalovich D., et al.,
Astron. Lett., {\bf 30}, 382 (2004)

\reference Romanova M. M., Ustyugova G. V., Koldoba A. V., Lovelace R. V. E.,
\apj\, {\bf 610}, 920 (2004)

\reference Santangelo A., Segreto A., Giarrusso, S. \etal,
\apj\, {\bf 523}, L85 (1999)

\reference Tamura K., Tsunemi H., Kitamoto S. \etal,
\apj\, {\bf 389}, 676 (1992)

\reference Tsygankov S., Lutovinov A., Churazov E., Sunyaev R.,
\mnras\ {\bf 371}, 19 (2006)

\reference White N., Swank J., Holt S.,
\apj\, {\bf 270}, 771 (1983)

\reference Wheaton W. A., Doty J. P., Primini F. A. \etal,  \nat\ {\bf 282}, 240 (1979)

\reference Winkler C., Courvoisier T.J.-L., Di Cocco G. \etal,  \aap\ {\bf 411}, L1 (2003)



\end{references}
\end{document}